\documentclass[conference]{IEEEtran}
\IEEEoverridecommandlockouts
\usepackage{cite}
\usepackage{amsmath,amssymb,amsfonts}
\usepackage{algorithm} 
\usepackage{tabularx}
\usepackage{subcaption}
\usepackage{algpseudocode}
\usepackage{graphicx}
\usepackage{textcomp}
\usepackage[table]{xcolor}
\usepackage{xcolor}
\def\BibTeX{{\rm B\kern-.05em{\sc i\kern-.025em b}\kern-.08em
    T\kern-.1667em\lower.7ex\hbox{E}\kern-.125emX}}

\newcommand{\LOOPLCP}{$\mathcal{LOOP-LC}\hspace{.1cm}2.0$}
\begin{document}

\title{Toward Rapid, Optimal, and Feasible Power Dispatch through Generalized Neural Mapping
%: A Machine Learning-Driven Methodology
}

\author{\IEEEauthorblockN{Meiyi Li$^1$,  Javad Mohammadi$^1$}
\IEEEauthorblockA{\textit{$^1$Department of Civil, Architectural, and Environmental Engineering, The University of Texas at Austin}\\
meiyil@utexas.edu, javadm@utexas.edu}
}

\maketitle
\begin{abstract}
The evolution towards a more distributed and interconnected grid necessitates large-scale decision-making within strict temporal constraints. Machine learning (ML) paradigms have demonstrated significant potential in improving the efficacy of optimization processes. However, the feasibility of solutions derived from ML models continues to pose challenges. It's imperative that ML models produce solutions
that are attainable and realistic within the given system constraints of power systems. To address the feasibility issue and expedite the solution search process, we proposed \LOOPLCP~(Learning to Optimize the Optimization Process with Linear Constraints version 2.0) as a learning-based approach for solving the power dispatch problem. A notable advantage of the \LOOPLCP~framework is its ability to ensure near-optimality and strict feasibility of solutions without depending on computationally intensive post-processing procedures, thus eliminating the need for iterative processes.
At the heart of the \LOOPLCP~model lies the newly proposed generalized gauge map method, capable of mapping any infeasible solution to a feasible point within the linearly-constrained domain. The proposed generalized gauge map method improves the traditional gauge map by exhibiting reduced sensitivity to input variances while increasing search speeds significantly. 
Utilizing the IEEE-200 test case as a benchmark, we demonstrate the effectiveness of the \LOOPLCP~methodology, confirming its superior performance in terms of training speed, computational time, optimality, and solution feasibility compared to existing methodologies.
\end{abstract}


\section{Introduction}

The ongoing evolution of the electrical grid is fuelled by increased incorporation of decentralized generation, distributed storage, and advancements in communication and sensing technologies. This evolution is also shaped by climate change concerns, the need for resilience, and ongoing electrification trends, resulting in a grid that is increasingly distributed and interconnected \cite{mohammadi2023towards,mohammadi2023empowering}. This transition demands decisions to be executed on a large scale within a narrow time frame.

Traditional optimization solvers, employing iterative algorithms, face challenges of extended calculation times, limiting their application in time-critical applications. Recent shifts towards machine learning (ML) aim to overcome such issues, enhancing optimization efficiency, particularly in power dispatch problems. The growing focus is largely on neural approximators to understand the relationship between varying input setups and their optimal solutions. Numerous studies \cite{tabas2022computationally,fioretto2020predicting,pan2022deepopf,wang2022fast,li2023learning} suggest that neural networks can significantly speed up the online search process, reducing the iterations required to find optimal solutions. Additionally, the repetitive nature of power dispatch problems generates abundant historical data. When leveraged through ML, allows for offline computational, which boosts the efficiency of real-time operations.

A notable challenge in applying ML methods to solve power dispatch problems lies in ensuring feasibility, i.e., adhering to the physical and engineering constraints that govern power flows and guarantee reliable operations. This implies that ML models should produce solutions that are attainable and realistic within the given system constraints, which is critical for maintaining stable and reliable power system operations.

Several strategies have been proposed to address the feasibility issue. Incorporating a penalty term to constrain the output of neural approximators is a straightforward tactic \cite{fioretto2020predicting,pan2022deepopf,wang2022fast}. Nonetheless, penalty-based methods only offer a soft boundary on the output as infeasibility is merely penalized, not eliminated. An alternative proposition involves modifying the constraint set that the ML model should learn, making it easier to learn. These endeavors include identifying a subset of active constraints \cite{hasan2021hybrid}, approximating the constraint set \cite{liang2023low}, or shrinking the constraint set \cite{tabas2022computationally}. However, these methods may result in infeasible or overly conservative solutions due to inappropriate constraint set reconstruction.

To ensure feasibility, projection techniques can be employed for infeasible solutions; however, solving the projection problem via an optimization solver \cite{zhao2020deepopf+} or initiating an additional iterative process \cite{donti2021dc3} proves to be computationally inefficient in real-time scenarios. Unlike these methods, our preceding work \cite{li2023learning} leverages gauge map functions to achieve hard feasibility of linear constraints. The gauge map method operates in a feed-forward manner and removes iterations entirely, which notably diminishes computation time and overhead. Nonetheless, the gauge map method forms a one-to-one mapping by rescaling centered at a specified interior point, which may lead to uneven distribution issues and undesired sensitivity regarding the interior points. Such challenges could slow down the search speed during offline model training.

In this paper, we introduce \LOOPLCP~(Learning to Optimize the Optimization Process with Linear Constraints version 2.0) as a learning-driven approach to address the power dispatch problem. A notable advantage of the proposed \LOOPLCP~method is its ability to ensure near-optimality and hard feasibility of solutions without the need for computationally demanding post-processing procedures, thereby eliminating iterative processes. This is achieved as \LOOPLCP~incorporates a newly proposed generalized gauge map method capable of converting any infeasible solution to a feasible point within the linearly-constrained set. Compared to the traditional gauge map method \cite{li2023learning}, the proposed generalized gauge map resolves unevenly-distribution issue and displays reduced sensitivity to input. Hence, the \LOOPLCP~approach with the generalized gauge map significantly enhances the search speeds. Through IEEE-200 test case, we demonstrate the effectiveness of the \LOOPLCP~method, confirming its superior performance concerning training speed, computational time, optimality, and solution feasibility when compared to existing methodologies.

\section{Problem Formulation}

In this section, we introduce the notations and problem
formulations. As the basic form of power system energy supply-demand problem,  power dispatch optimization seeks to find the most cost-effective power production to meet the end-users' needs. This can be formulated as,

\begin{subequations}
\label{power dispatch problem}
\begin{gather}
\texttt{cost function: }\min f(\mathbf{u})\label{objective function}\\
\texttt{supply-demand balance: }{\mathbf{1}}^{\texttt{T}}\mathbf{u}={\mathbf{1}}^{\texttt{T}}\mathbf{x}\label{balance} \\
\texttt{generation limit: }\mathbf{u}_\texttt{min}\leq \mathbf{u}\leq \mathbf{u}_\texttt{max}\label{generation_limit}  
\end{gather}
\end{subequations}
where $\mathbf{u}$ represents the vector of the electric power production, and $f(\mathbf{u})$ denotes the associated overall cost function. Equations \eqref{balance} and \eqref{generation_limit} establish element-wise equality and inequality relations, where  $\mathbf{x}$ refers to the electric demand vector of all modes, with the  term `node' referring to an end-user or consumer, $\mathbf{1}$ denotes an all-one vector, $\mathbf{u}_\texttt{min}$ and $\mathbf{u}_\texttt{max}$ denote physical limitations on power generation.

To streamline the notation, we will denote the optimal solution and constraint set of problem \eqref{power dispatch problem} as $\mathbf{u}^*$ and $\mathcal{S}$, respectively.
%, with $\mathcal{S}$ being non-empty. 
An intuitive solution to problem \eqref{power dispatch problem} is represented as $\mathbf{u}_\texttt{o}$, expressed as follows:
\begin{align}
    \mathbf{u}_\texttt{o}= \mathbf{u}_\texttt{min}+\frac{{\mathbf{1}}^{\texttt{T}}\mathbf{x}-{\mathbf{1}}^{\texttt{T}}\mathbf{u}_\texttt{min}}{{\mathbf{1}}^{\texttt{T}}\mathbf{u}_\texttt{max}-{\mathbf{1}}^{\texttt{T}}\mathbf{u}_\texttt{min}}(\mathbf{u}_\texttt{max}-\mathbf{u}_\texttt{min})\label{int}
\end{align}

The power dispatch problem \eqref{power dispatch problem}, is typically resolved repeatedly. It may be solved hourly, utilizing forecasted electric demand for the upcoming hour's preparation, or every 5 minutes in accordance with the production schedule to maintain real-time power equilibrium within a designated look-ahead horizon. Instead of solving the optimization problem \eqref{power dispatch problem} by an iterative solver, we aim to approximate this repeated process with a neural network approximator, $\xi$. The approximator directly maps the input to the optimal solution in a single feed-forward. We use $\blacktriangle$ to denote the resulting prediction, and ideally,  $\mathbf{u}^\blacktriangle $ will be close to optimal solution $\mathbf{u}^*$ of  \eqref{power dispatch problem}.  
\begin{align}
    \mathbf{u}^\blacktriangle={\xi}(\mathbf{x},\mathbf{u}_\texttt{o})\label{eq:nn}
\end{align}

\begin{figure}[htbp]
\centering
\setlength{\abovecaptionskip}{0.cm}
\includegraphics[width=0.8\columnwidth]{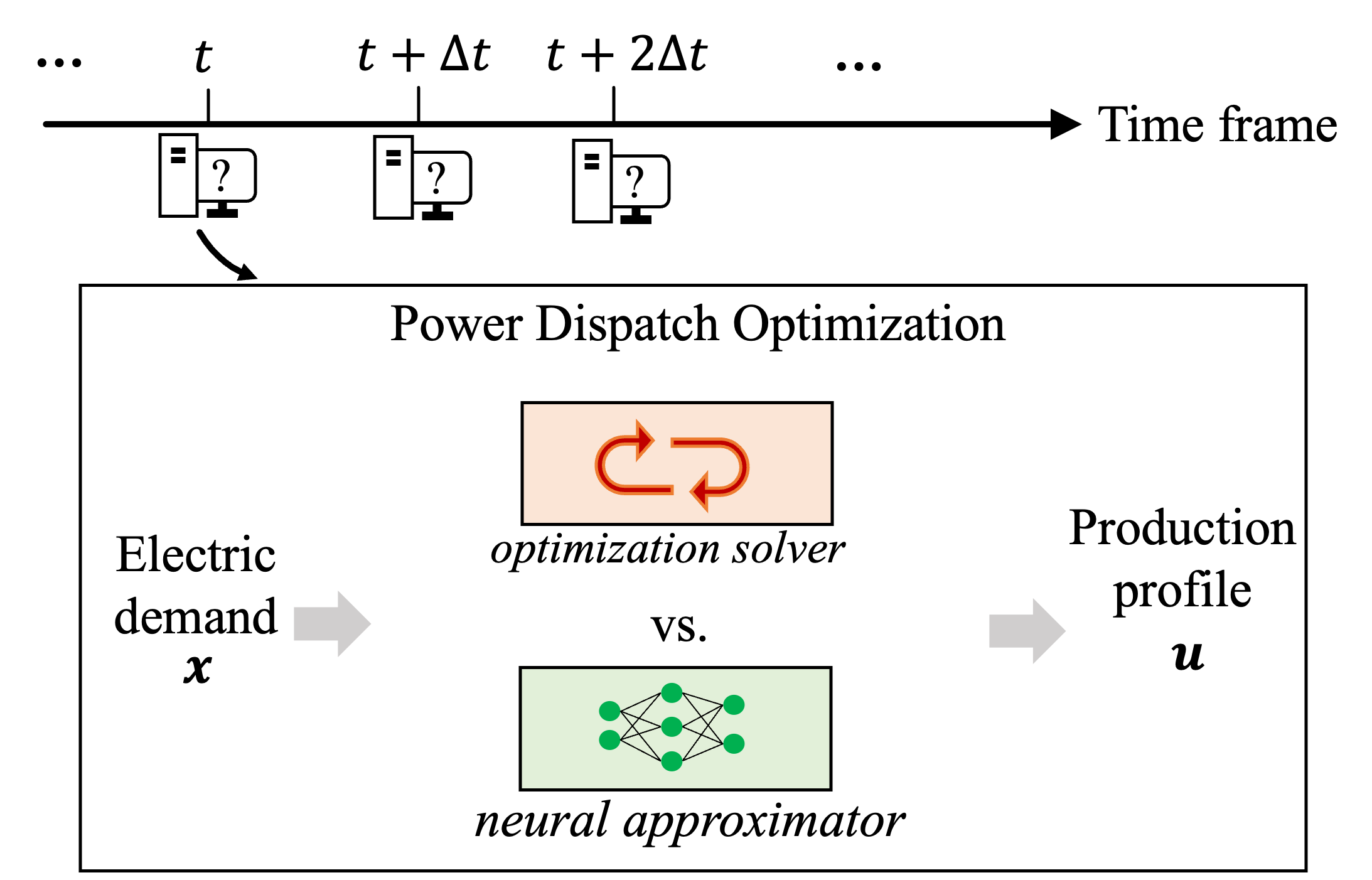}
\caption{We propose a \LOOPLCP~model as a Neural Approximator to replace iterative solvers for power dispatch optimization problem.}\centering
\label{f:basic}
\end{figure}

A significant challenge in utilizing ML methods for solving power dispatch problems is ensuring feasibility, i.e., $\mathbf{u}^\blacktriangle$ adheres to the constraint set $\mathcal{S}$. In the subsequent section, we will present the detailed \LOOPLCP~method to train and obtain $\xi$. Central to the \LOOPLCP~model is a newly proposed generalized gauge map method, which can ensure hard feasibility and eliminate iterations for the post-projection procedure altogether.

\begin{figure*}[htbp]
\centering
\setlength{\abovecaptionskip}{0.2cm}
\includegraphics[width=1.95\columnwidth]{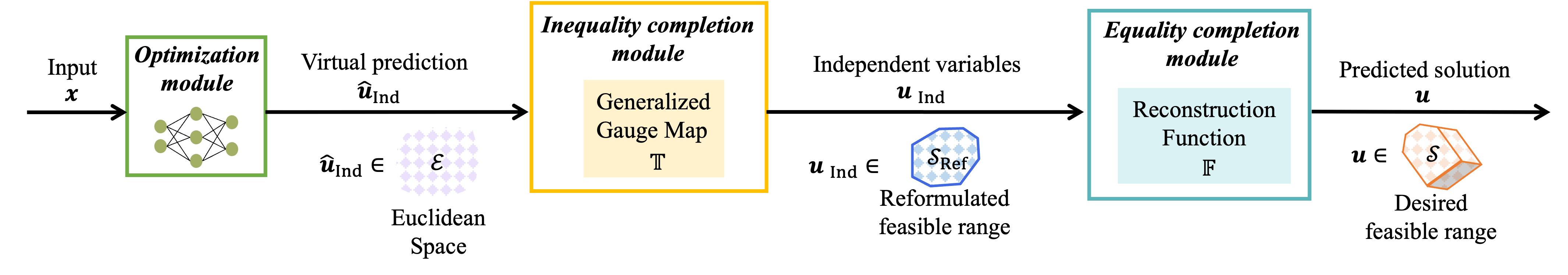}
\caption{The \LOOPLCP~framework is composed of (i) optimization, (ii) inequality completion, and (iii) equality completion modules.} 
\label{f:module}
\end{figure*}

\section{The proposed \LOOPLCP~method}

% \subsection{Overview of \LOOPLCP~method}

This section outlines our proposed \LOOPLCP~model, which utilizes ML to solve optimization problems with linear constraints. The \LOOPLCP~framework comprises of (i) optimization, (ii) inequality completion, and (iii) equality completion modules. As illustrated in Fig. \ref{f:module}, the modular design of \LOOPLCP~aims to yield a high-quality (near-optimal) feasible solution. The core of the \LOOPLCP~model sits in the inequality completion module, where a newly proposed generalized gauge map method is employed to map any infeasible solutions to the desired feasible range. In what follows, a comprehensive elaboration of the modules within the \LOOPLCP~model is provided.

% The application of the LOOP − LC
% model to the proposed method (as shown in Fig.3) is explained
% belowto incorporate an ML method to accelerate the ADMM algorithm. Instead of solving the ${N}_{\texttt{A}}$ local optimization problems \eqref{eq:local_suboptimization_simple} by an iterative solver, we will train ${N}_{\texttt{A}}$ neural networks $\xi_{\mathbf{w}^i},i\in \mathcal{N}_{\texttt{A}}$ to directly map the input to the convergence value of local variables in a single feed-forward.
% Generalized Gauge Map
% In this section, we will present a refined version of the $\mathcal{LOOP-LC}$ model \cite{li2023learning}. The refined method is capable of rescaling
% any infeasible solution and mapping it to a feasible point
% within the linearly-constrained set. Compared to the $\mathcal{LOOP-LC}$ model, the refined method exhibits reduced sensitivity to
% the input interior point, resulting in substantially accelerated
% search speeds. We will first overview the $\mathcal{LOOP-LC}$ method and then present the details of our refined method.

% \subsection{Overview of $\mathcal{LOOP-LC}$ method}
% The $\mathcal{LOOP-LC}$ model learns to solve optimization problems with hard linear constraints by applying variable elimination and gauge mapping.

\subsection{Equality Completion
Module}
The equality completion
module in the \LOOPLCP~model achieves feasibility of equality constraints by variable elimination.  According to equality constraints \eqref{balance}, we divide the variables $\mathbf{u}$ into two parts where \textit{dependent variables} $\mathbf{u}_{\texttt{Dep}}$ are determined by \textit{independent variables} $\mathbf{u}_{\texttt{Ind}}$. For example, consider the first element in $\mathbf{u}$  as dependent variables relying on other elements in $\mathbf{u}$ with the relationship denoted as $\mathbb{F}$:

\begin{align}
\mathbf{u}_{\texttt{Dep}}=\mathbb{F}(\mathbf{u}_{\texttt{Ind}})={\mathbf{1}}^{\texttt{T}}\mathbf{x}-{\mathbf{1}}^{\texttt{T}}\mathbf{u}_{\texttt{Ind}}
    \label{eq:F}
\end{align}

% \mathbf{u}=\begin{bmatrix}
% \mathbf{u}_{\texttt{Dep}}\\ 
% \mathbf{u}_{\texttt{Ind}}
% \end{bmatrix}=\begin{bmatrix}
% \mathbb{F}(\mathbf{u}_{\texttt{Ind}})\\ 
% \mathbf{u}_{\texttt{Ind}}
% \end{bmatrix}=\begin{bmatrix}
% {\mathbf{1}}^{\texttt{T}}\mathbf{x}-{\mathbf{1}}^{\texttt{T}}\mathbf{u}_{\texttt{Ind}}\\ 
% \mathbf{u}_{\texttt{Ind}}
% \end{bmatrix}

Therefore, given $\mathbf{u}_{\texttt{Ind}}$, the equality completion module applies $\mathbb{F}$ to produce a full-size $\mathbf{u}=[\mathbf{u}_{\texttt{Dep}},\mathbf{u}_{\texttt{Ind}}]$. The definition of $\mathbb{F}$ in \eqref{eq:F} ensures  
the produced $ \mathbf{u}$ satisfies \eqref{balance}.

\subsection{Inequality Completion
Module}
\subsubsection{Functional Requirements of Inequality Completion
Module}Let's first specify the functional requirement of inequality completion
module. 
Take the relationship $\mathbb{F}$ into the definition of 
$\mathcal{S}$ and replace $\mathbf{u}_{\texttt{Dep}}$, we could reformulate the problem \eqref{power dispatch problem} to a reduced-size problem whose variable is $\mathbf{u}_{\texttt{Ind}}$. 

The reformulated constraints set  $\mathcal{S}_{\texttt{Ref}}$ is presented as \eqref{sref}. 

\begin{align}
    \mathcal{S}^{\texttt{Ref}}=\left \{ \mathbf{u}^{\texttt{Indep}}|\mathbf{A}\mathbf{u}^{\texttt{Indep}}\leq \mathbf{B}\mathbf{x}+\mathbf{b} \right \} \label{sref}
\end{align}
where $$\mathbf{A}=\begin{bmatrix}
-{\mathbf{1}}^{\texttt{T}}\\ 
\mathbf{I}\\ 
{\mathbf{1}}^{\texttt{T}}\\ 
-\mathbf{I} 
\end{bmatrix},\mathbf{B}=\begin{bmatrix}
-{\mathbf{1}}^{\texttt{T}}\\ 
{\mathbf{0}}^{\texttt{T}}\\
...\\ 
{\mathbf{1}}^{\texttt{T}}\\ 
{\mathbf{0}}^{\texttt{T}}\\
... 
\end{bmatrix},\mathbf{b}=\begin{bmatrix}
\mathbf{u}_\texttt{max}\\
\mathbf{u}_\texttt{min}
\end{bmatrix}$$

Here, $\mathbf{0}$ denotes an all-one vector, $\mathbf{I}$ denotes a unit matrix. The functional requirement of inequality completion
module is to make sure the module's output $\mathbf{u}_{\texttt{Ind}}\in\mathcal{S}_{\texttt{Ref}}$.

\subsubsection{Traditional gauge map method and its issues}
To derive the inequality completion module's output $\mathbf{u}_{\texttt{Ind}}\in\mathcal{S}_{\texttt{Ref}}$, we could use the gauge map method presented in \cite{li2023learning}. Note, $\mathcal{S}_{\texttt{Ref}}$ is a set defined with linear constraints, i.e., a polytope. The gauge map method could establish a one-to-one mapping between points within $\ell_\infty$-norm unit ball $\mathcal{B}$ and points within the polytope. According to \cite{li2023learning}, given any $\mathbf{\hat{u}}_{\texttt{Ind}}\in \mathcal{B}$, the gauge mapping is a known function with a closed form of:

\begin{align}   \mathbf{u}_{\texttt{Ind}}=\frac{\psi_{\mathcal{B}}(\mathbf{\hat{u}}_{\texttt{Ind}})}{\psi_{\mathcal{S}_{\texttt{Ref0}}}(\mathbf{\hat{u}}_{\texttt{Ind}})}\mathbf{\hat{u}}_{\texttt{Ind}}+\mathbf{u}_{\texttt{Ind,0}}\label{eq:T}
\end{align}
where $\psi_{\mathcal{B}}(\mathbf{\hat{u}}_{\texttt{Ind}})$ is the Minkowski function value of $\mathbf{\hat{u}}_{\texttt{Ind}}$ on set $\mathcal{B}$, defined as \eqref{psi_b}. Note, $\psi_{\mathcal{S}_{\texttt{Ref0}}}(\mathbf{\hat{u}}_{\texttt{Ind}})$ is the Minkowski function value of $\mathbf{\hat{u}}_{\texttt{Ind}}$ on set $\psi_{\mathcal{S}_{\texttt{Ref0}}}$, defined as \eqref{psi_s}. 
$\mathbf{u}_{\texttt{Ind,0}}$ 
is the interior point of $\mathcal{S}_{\texttt{Ref}}$ as defined in \eqref{int}. Also, $\mathcal{S}_{\texttt{Ref0}}=\left \{ \mathbf{\bar{u}}_{\texttt{Ind}}| \left (\mathbf{u}_{\texttt{Ind,0}}+\mathbf{\bar{u}}_{\texttt{Ind}}  \right )\in \mathcal{S}_{\texttt{Ref}}\right \}$ is a shifted set. In the definition of Minkowski function \eqref{psi_b} and \eqref{psi_s}, the superscript $r$ denotes the $r$th row in a vector (or matrix).

\begin{align}
    \psi_{\mathcal{B}}(\mathbf{\hat{u}}_{\texttt{Ind}})=\underset{r}{\max}\left \{ \left | \mathbf{\hat{u}}_{\texttt{Ind}}^r \right | \right \}\label{psi_b}\\
    \psi_{\mathcal{S}_{\texttt{Ref0}}}(\mathbf{\hat{u}}_{\texttt{Ind}})=\underset{r}{\max}\{\frac{\mathbf{A}^r\mathbf{\hat{u}}_{\texttt{Ind}}}{{(-\mathbf{A}\mathbf{u}_{\texttt{Ind,0}}+\mathbf{B}\mathbf{x}+\mathbf{b})}^r}\} \label{psi_s}
\end{align}

The gauge map method ensures feasibility within a desired polytope by rescaling centered at a given interior point. 
Therefore, given an interior point $\mathbf{u}_{\texttt{Ind,0}}$ and a virtual prediction $\mathbf{\hat{u}}_{\texttt{Ind}}\in \mathcal{B}$, the gauge map as \eqref{eq:T} could achieve the functional requirement of inequality completion module of ensuring $\mathbf{u}_{\texttt{Ind}}\in \mathcal{S}_{\texttt{Ref}}$.

\begin{figure}[htbp]
\centering
\setlength{\abovecaptionskip}{0.2cm}
\includegraphics[width=1\columnwidth]{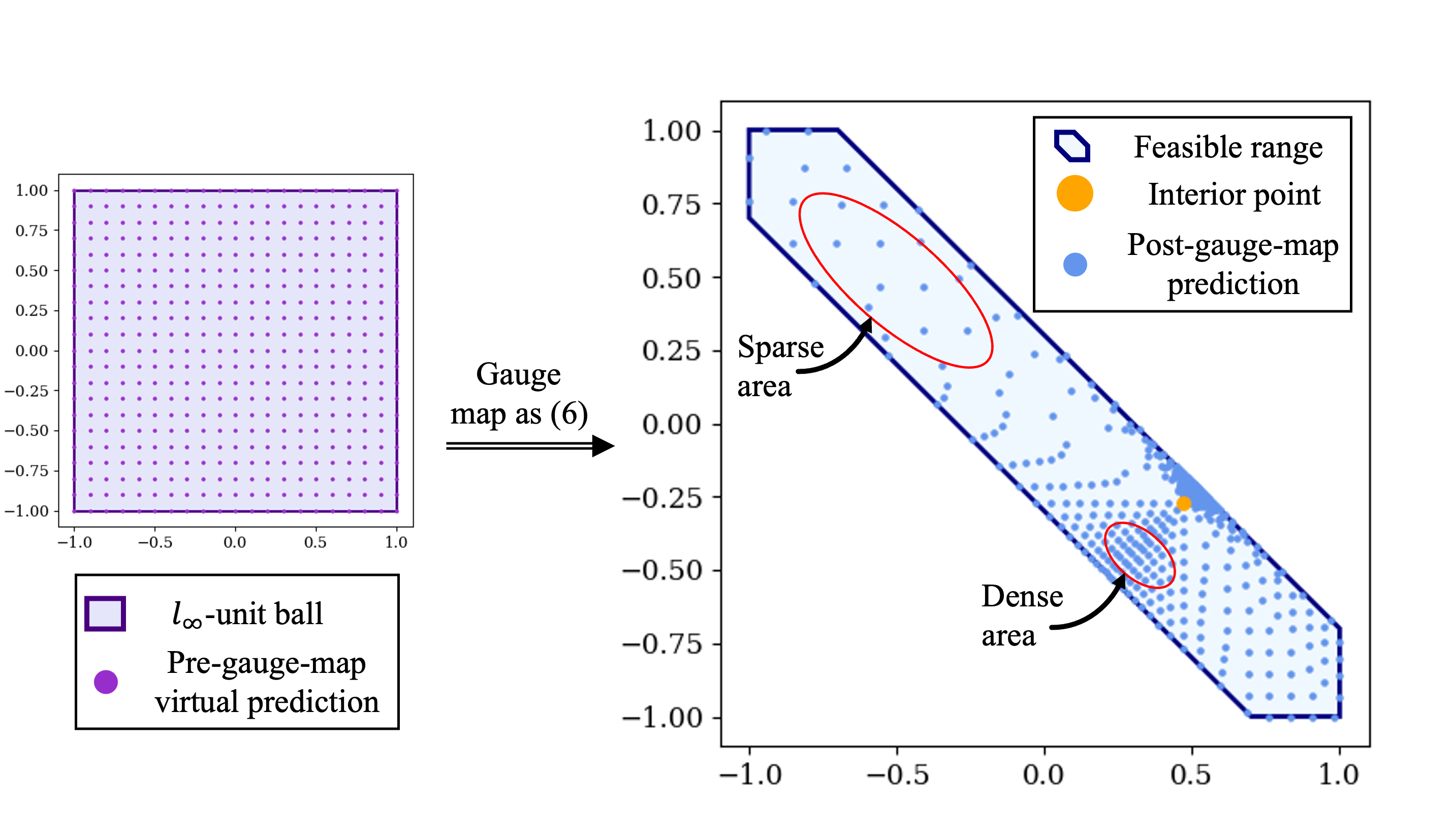}
\caption{A two-dimension example showing unevenly distributed predictions $\mathbf{u}_{\texttt{Ind}}$ caused by the rescaling through the traditional gauge map method presented in \eqref{eq:T}.} 
\label{f:scale}
\end{figure}

However, given evenly distributed virtual predictions $\mathbf{\hat{u}}_{\texttt{Ind}}$ in the unit box, the gauge map will output unevenly-distributed predictions in $\mathcal{S}_{\texttt{Ref}}$, as shown in Fig \ref{f:scale}. The uneven distribution will challenge the search process (using gradient descent for training) of neural networks. For example, the neural network might get trapped in the dense area because even significant variations in $\mathbf{\hat{u}}_{\texttt{Ind}}$ will result in similar predictions $\mathbf{u}_{\texttt{Ind}}$. Whereas in the sparse area, it is easy to miss the correct prediction $\mathbf{u}_{\texttt{Ind}}$ because slight change of the neural network output $\mathbf{\hat{u}}_{\texttt{Ind}}$ leads to predictions $\mathbf{u}_{\texttt{Ind}}$ far from each other.

\subsubsection{Proposed generalized gauge map}

\begin{figure}[htbp]
\begin{subfigure}[t]{0.47\columnwidth}
    \includegraphics[width=\columnwidth]{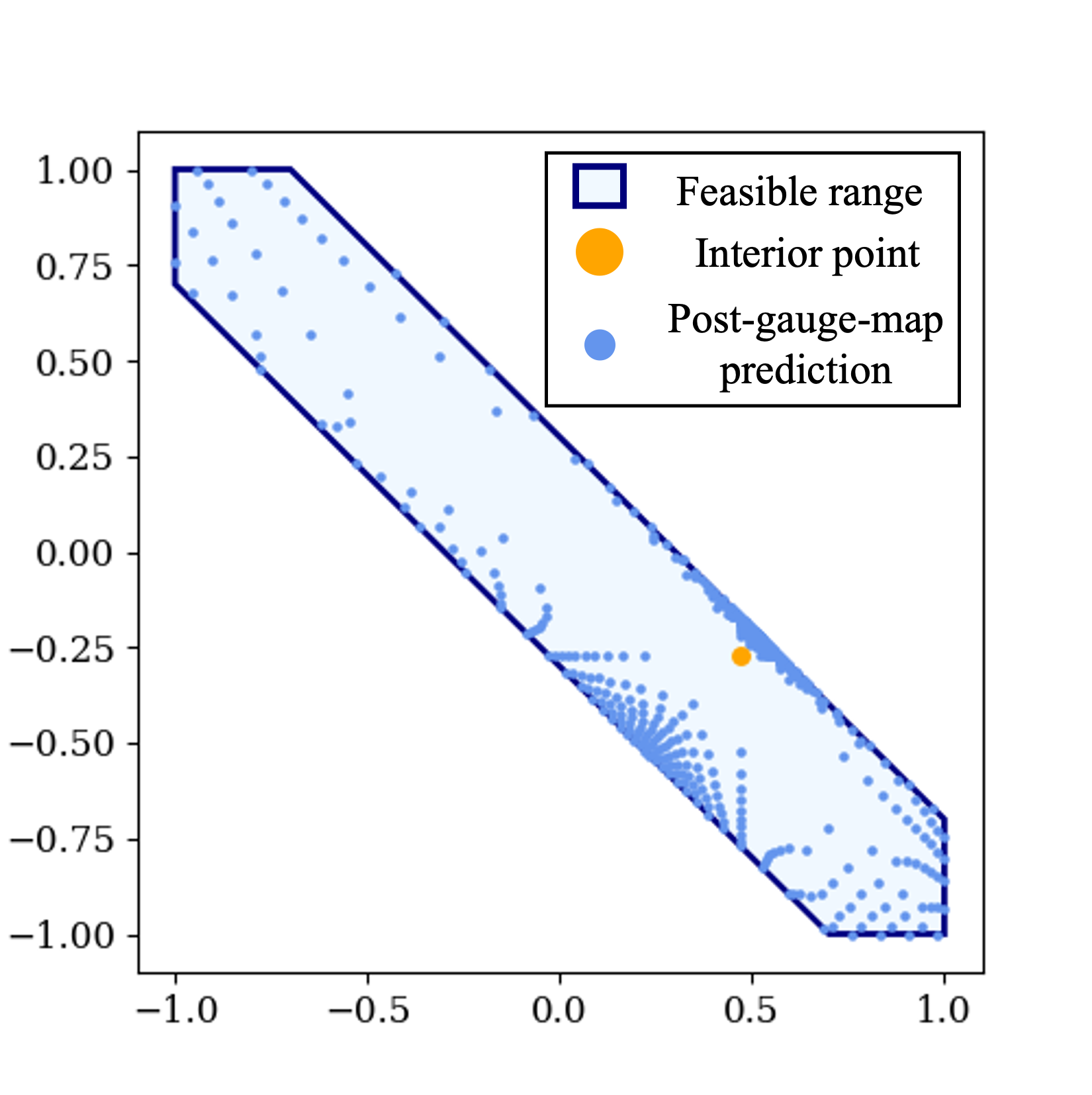}
    \caption{substitute $\varphi_{\mathcal{B}}^{0.3}$}
\end{subfigure}\hspace{\fill} % maximize horizontal separation
\begin{subfigure}[t]{0.47\columnwidth}
    \includegraphics[width=\linewidth]{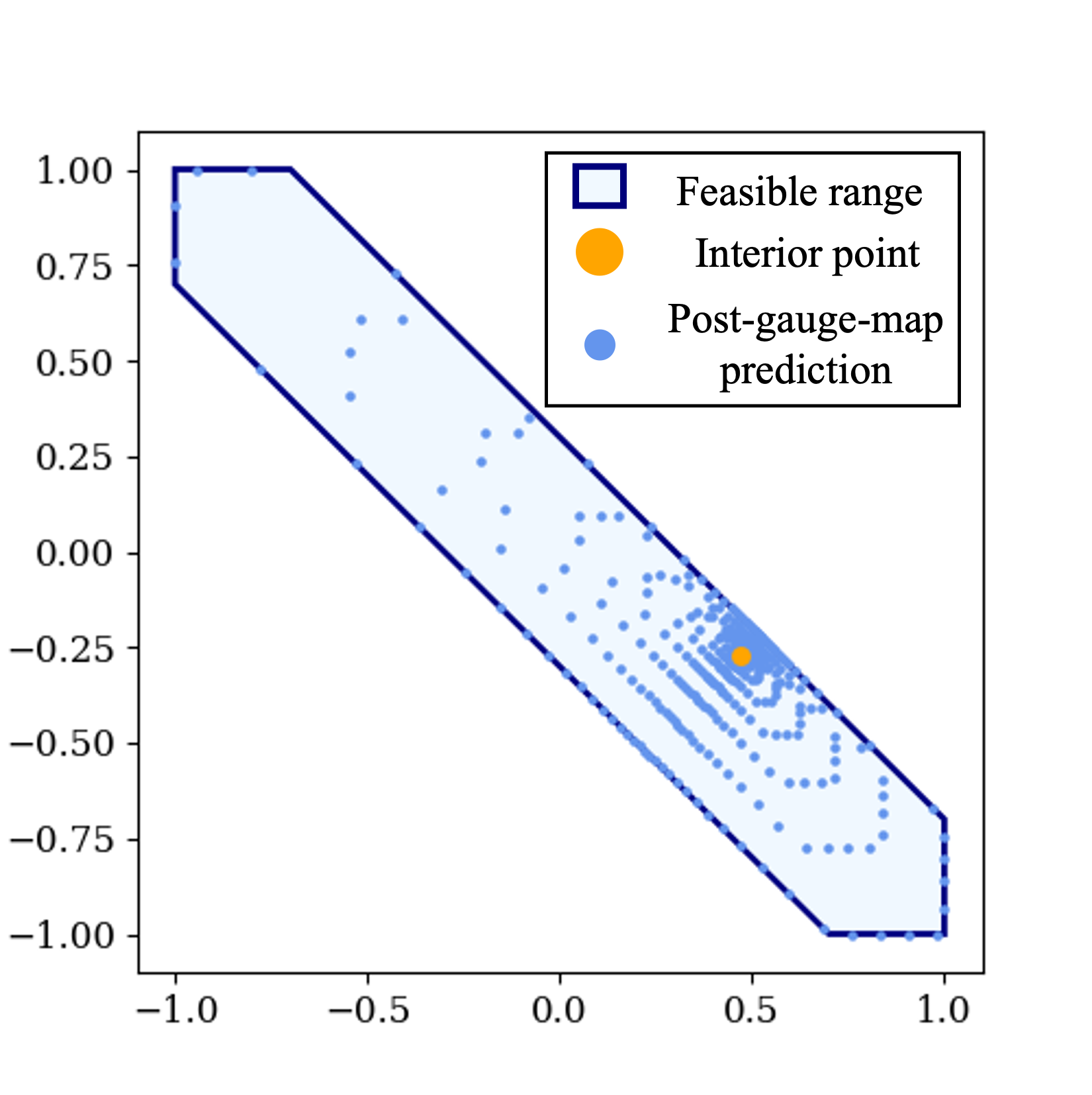}
    \caption{substitute $\varphi_{\mathcal{B}}^{3.5}$.}   
\end{subfigure}
\bigskip % more vertical separation
\begin{subfigure}[t]{0.47\columnwidth}
    \includegraphics[width=\linewidth]{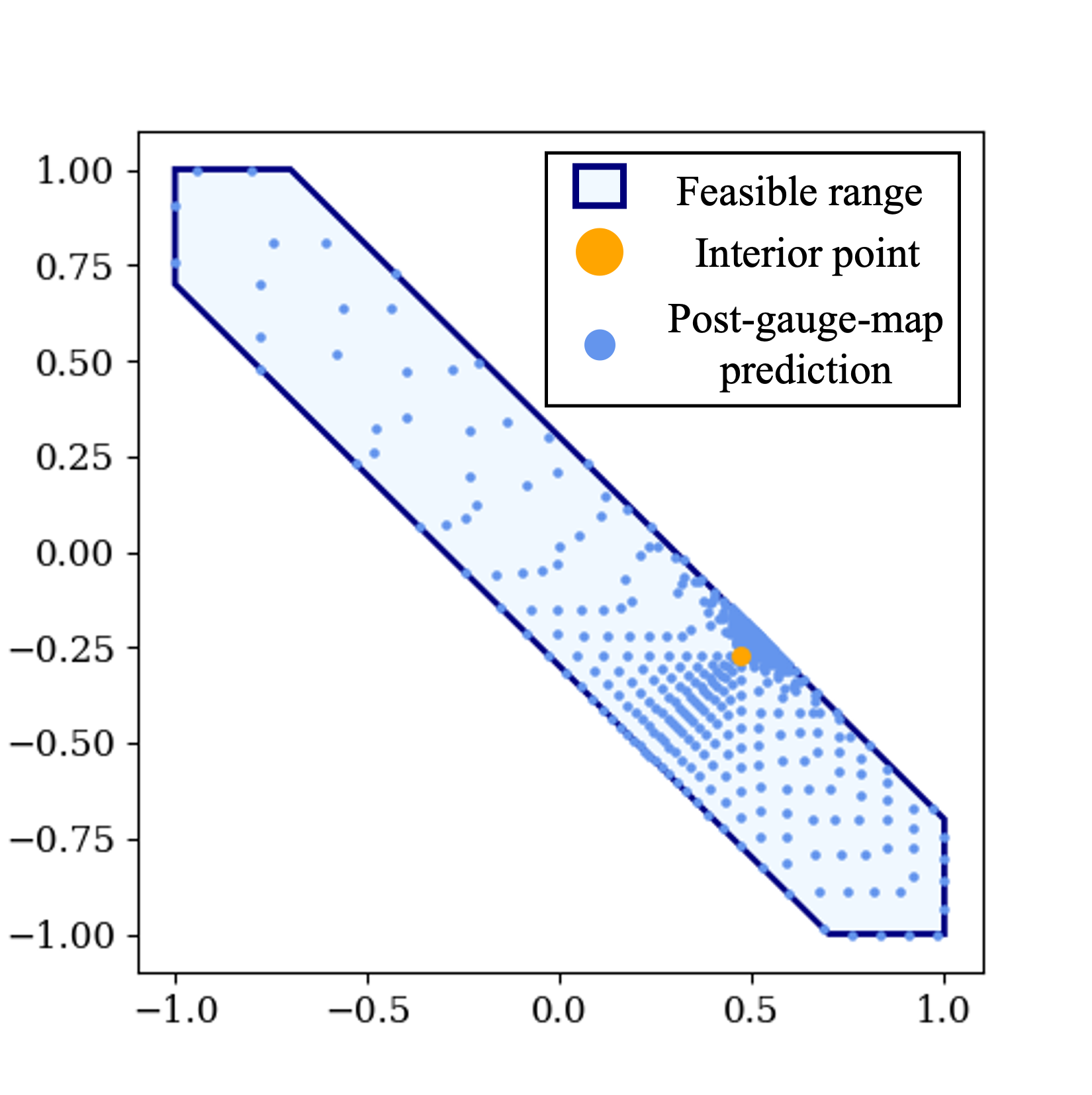}
    \caption{substitute $\frac{e^{\varphi_{\mathcal{B}}}-1}{(e-1)}$.}    
\end{subfigure}\hspace{\fill} % maximize horizontal separation
\begin{subfigure}[t]{0.47\columnwidth}
    \includegraphics[width=\linewidth]{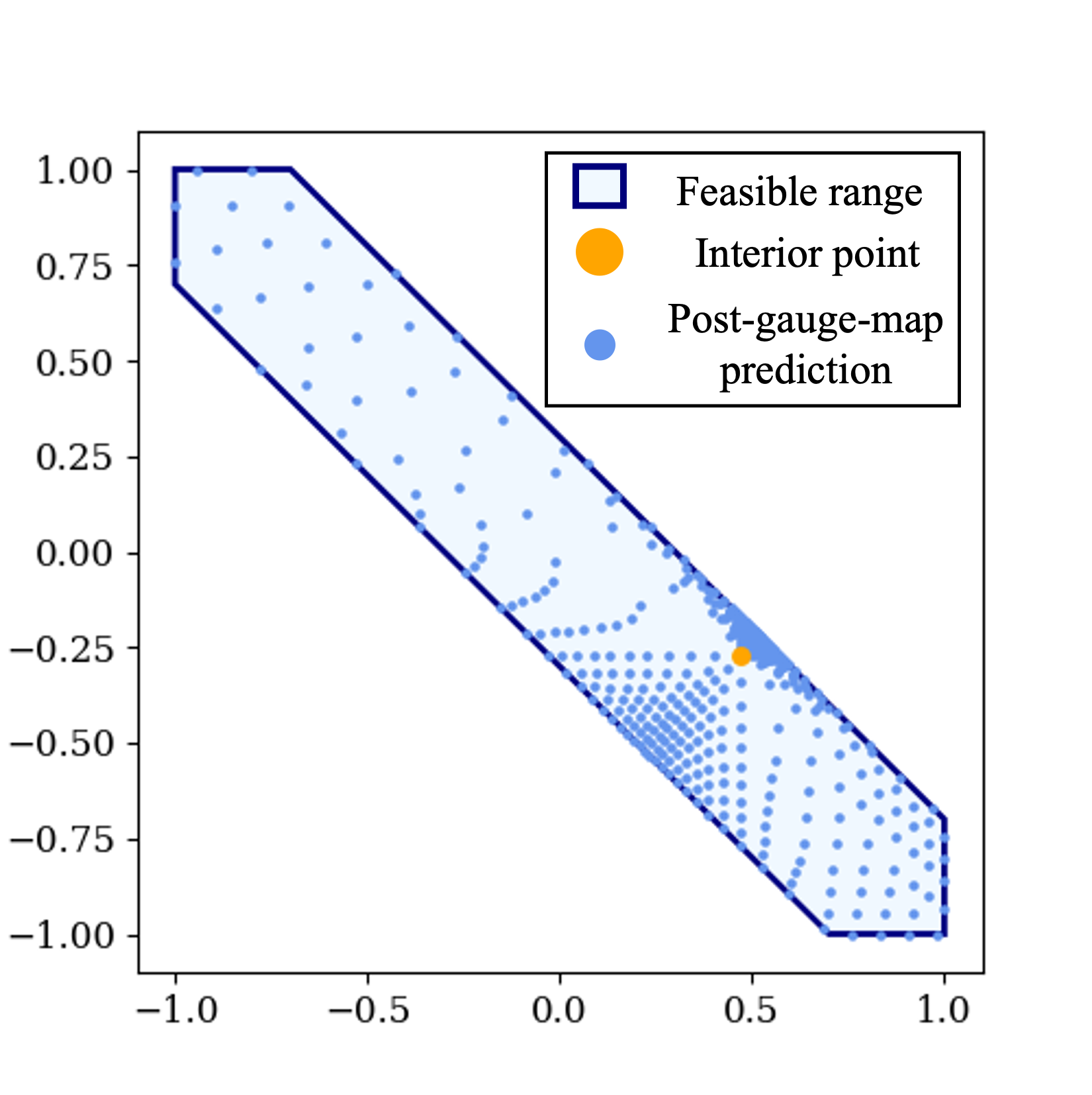}
    \caption{substitute $\frac{\log({\varphi_{\mathcal{B}}}+1)}{(\log2)}
$.}  
\end{subfigure}
\caption{Distribution of $\mathbf{u}_{\texttt{Ind}}$ using various  gauge map functions where we replace $\varphi_{\mathcal{B}}$. The predictions $\mathbf{u}_{\texttt{Ind}}$ will be densely distributed around the interior point when we substitute $p$th power($p\geq1$) of $\varphi_{\mathcal{B}}$ or its exponential form. Whereas $\mathbf{u}_{\texttt{Ind}}$ will be densely distributed around the boundaries using the log form or $p<1$. }
\label{f:some_new}
\end{figure}

Note that the traditional gauge map function achieves rescaling by multiplying the Minkowski distance function in set, i.e., $\mathcal{B}$. In fact, we could replace $\varphi_{\mathcal{B}}$ with its other forms as long as it still produces a continuous distance within $[0,1]$. For example, Fig.\ref{f:some_new} shows some of the other gauge map functions where we replace the $\varphi_{\mathcal{B}}$ with its $p$th power, its exponential form, and its log form.

Although the forms in Fig.\ref{f:some_new} could somewhat adjust the unevenly-distribution, the performance is still poor. To mitigate the unevenly-distribution issue and save training time, we propose a new generalized gauge map function as below:

\begin{align}
\mathbf{u}_{\texttt{Ind}}=\mathbb{T}(\mathbf{\hat{u}}_{\texttt{Ind}})=\frac{1}{\max\left \{ 1, \psi_{\mathcal{S}_{\texttt{Ref0}}}(\mathbf{\hat{u}}_{\texttt{Ind}})\right \}}\mathbf{\hat{u}}_{\texttt{Ind}}+\mathbf{u}_{\texttt{Ind,0}}\label{eq:new_gauge_map}
\end{align}

%     \mathbf{u}=\mathbb{T}({\mathbf{v}})
%     =\left\{\begin{matrix}
% \frac{\varphi_{\mathcal{\bar{S}}^{\texttt{Ref}}}(\mathbf{v})}{\varphi_{\mathcal{\bar{S}}^{\texttt{Ref}}}(\mathbf{v})}\mathbf{v}+\mathbf{u}_{\texttt{o}}=\mathbf{v}+\mathbf{u}_{\texttt{o}},\mathbf{v}\in \mathcal{\bar{S}}^{\texttt{Ref}}
% \\ \frac{1}{\varphi_{\mathcal{\bar{S}}^{\texttt{Ref}}}(\mathbf{v})}\mathbf{v}+\mathbf{u}_{\texttt{o}},~~~~~~~~~~\mathbf{v}\notin \mathcal{\bar{S}}^{\texttt{Ref}}
% \end{matrix}\right.

\begin{figure}[htbp]
\centering
\setlength{\abovecaptionskip}{0.2cm}
\includegraphics[width=0.5\columnwidth]{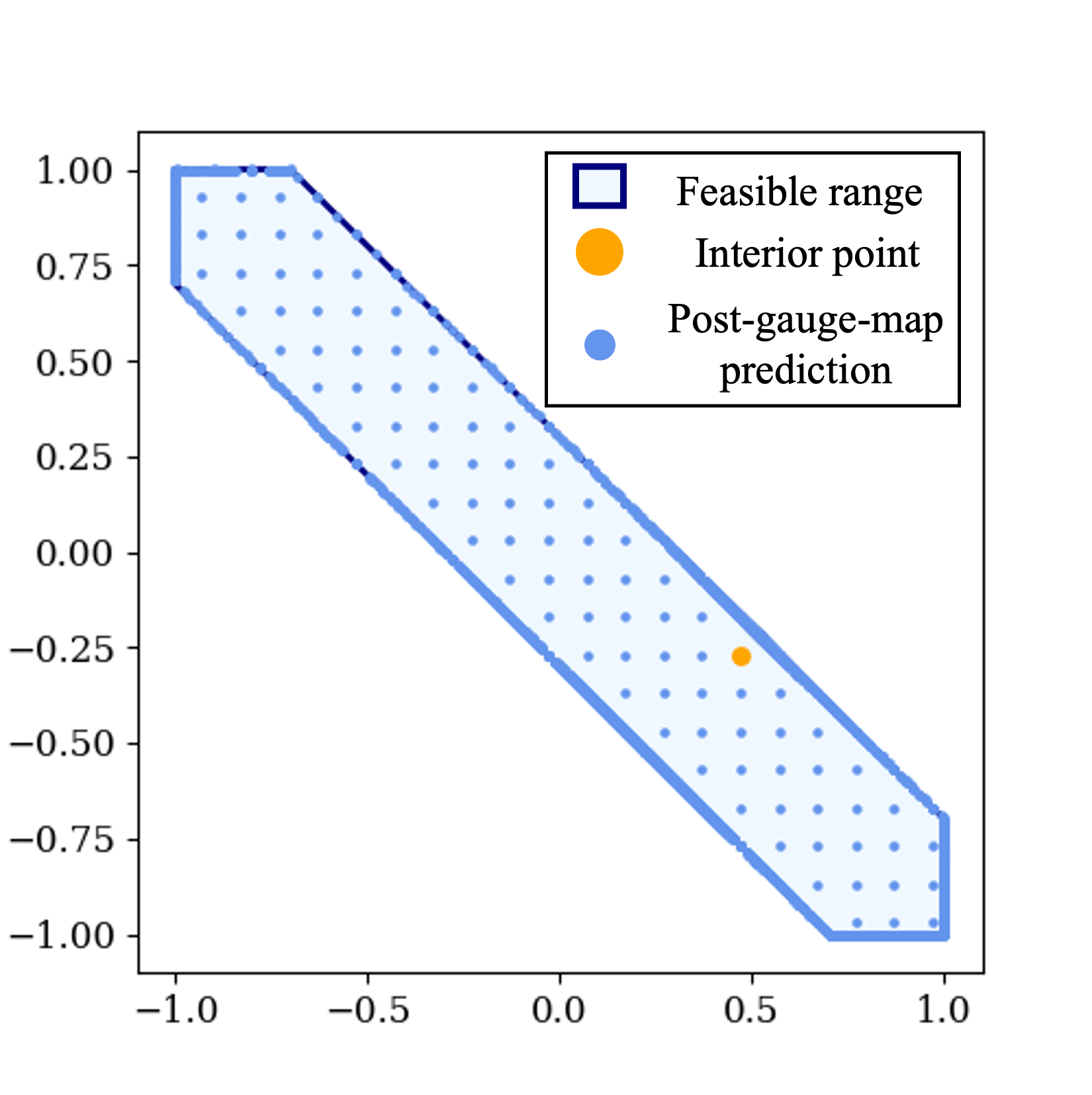}
\caption{The porposed generalized gauge map function will keep the virtual predictions as it is if they have already lie in the desired feasible range. Otherwise, the generalized gauge map function will rescale the virtual predictions to the boundary.}
\label{f:newg}
\end{figure}

The proposed generalized gauge map works as below:
\begin{itemize}
    \item when given a virtual prediction $\mathbf{\hat{u}}_{\texttt{Ind}}\in \mathcal{S}_{\texttt{Ref0}}$, $\psi_{\mathcal{S}_{\texttt{Ref0}}}(\mathbf{\hat{u}}_{\texttt{Ind}})\leq1$. The generalized gauge map \eqref{eq:new_gauge_map} reduces to $\mathbf{u}_{\texttt{Ind}}=\mathbf{\hat{u}}_{\texttt{Ind}}+\mathbf{u}_{\texttt{Ind,0}}$, which keeps the virtual prediction as it is and hence $\mathbf{u}_{\texttt{Ind}}\in \mathcal{S}_{\texttt{Ref}}$.
    \item While when the virtual prediction $\mathbf{\hat{u}}_{\texttt{Ind}}\notin\mathcal{S}_{\texttt{Ref0}}$, $\psi_{\mathcal{S}_{\texttt{Ref0}}}(\mathbf{\hat{u}}_{\texttt{Ind}})>1$. The generalized gauge map will transfer the infeasible point to the boundary as $\mathbf{u}_{\texttt{Ind}}=\frac{1}{ \psi_{\mathcal{S}_{\texttt{Ref0}}}(\mathbf{\hat{u}}_{\texttt{Ind}})}\mathbf{\hat{u}}_{\texttt{Ind}}+\mathbf{u}_{\texttt{Ind,0}}$.
\end{itemize}
As shown in Fig.\ref{f:newg}, the proposed generalized gauge map ensures  $\mathbf{u}_{\texttt{Ind}}\in \mathcal{S}_{\texttt{Ref}}$ given any $\mathbf{\hat{u}}_{\texttt{Ind}}$.

To sum up, the proposed generalized gauge map could rescale any infeasible solutions to the boundary of constraint set whereas feasible points remains unchanged. The proposed generalized gauge map function exhibits reduced sensitivity to
the given interior point and results in a substantially accelerated search process by solving the unevenly-distribution issue of traditional gauge maps. The proposed generalized gauge map function could be easily implemented in code format, and it allows backward propagation, which means it has the potential to be integrated into any differentiable ML models.

\subsection{Optimization Module}
The optimization module uses a neural network to learn a virtual prediction $\mathbf{\hat{u}}_{\texttt{Ind}}$.
Note that to make sure that any points in the feasible range could be reached out, the neural network in the optimization module should be able to produce virtual predictions $\mathbf{\hat{u}}_{\texttt{Ind}}\in \mathcal{E}$, where $\mathcal{E}$ is a superset of $\mathcal{S}_{\texttt{Ref0}}$, i.e., $\mathcal{S}_{\texttt{Ref0}}\subseteq  \mathcal{E}$. In this paper, we consider $\mathcal{E}$ is the Euclidean space $\mathcal{R}^{n_{\texttt{Ind}}}$, where $n_{\texttt{Ind}}$ is the dimension of $\mathbf{u}_{\texttt{Ind}}$.

The optimality module uses two training approaches
1) with a solver in the loop, and 2) without a solver in the loop, i.e., directly minimizing the objective function. In this paper, we consider using the distance between the prediction $\mathbf{u}^\blacktriangle$ and the optimal solution $\mathbf{u}^*$(calculated using commercial solvers) of problem \eqref{power dispatch problem} as the loss function, as \eqref{loss}. 

\begin{align}
    L=\frac{1}{N}\sum_{n=1}^{N} \left \| \mathbf{u}^{(n)\blacktriangle},\mathbf{u}^{(n)*} \right \|^2_2\label{loss}
\end{align}

\noindent where $N$ denotes the number of input data points, and $n$ denotes its index.

\section{Case study}
%Here, we present the \LOOPLCP~'s test results.
\subsection{Dataset and test configurations}
We use the publicly available IEEE 200-bus system data set, available via the MATPOWER \cite{zimmerman2010matpower}, as the seed information to
generate 200 data points (with a train/test ratio of 1:1). We consider a 10-percentage fluctuation of each load node.  

We compare \LOOPLCP~against the two learning-based optimization methods; (i) penalty  method \cite{liu2022teaching}, and (ii) the traditional gauge method \cite{li2023learning}, (iii) projection method \cite{zhao2020deepopf+}, as well as the well-known commercial solver GUROBI \cite{gurobi}. The penalty method adds a $\ell_2$-norm term to the loss function with penalty coefficient $\rho>0$. The traditional gauge map method utilize a mapping function as \eqref{eq:T}.

We use a fixed neural network architecture: fully connected with one hidden layer of size 64, including the rectified linear unit (ReLU) activation. An extra Tanh activation is added to the output layer for the traditional gauge map method.  Different hyperparameters $\rho$ were tuned between 1e-7 to 100 to maximize performance for penalty method.

\subsection{Results regarding offline training speed}
\begin{figure}[htbp]
\centering
\setlength{\abovecaptionskip}{0.2cm}
\includegraphics[width=1\columnwidth]{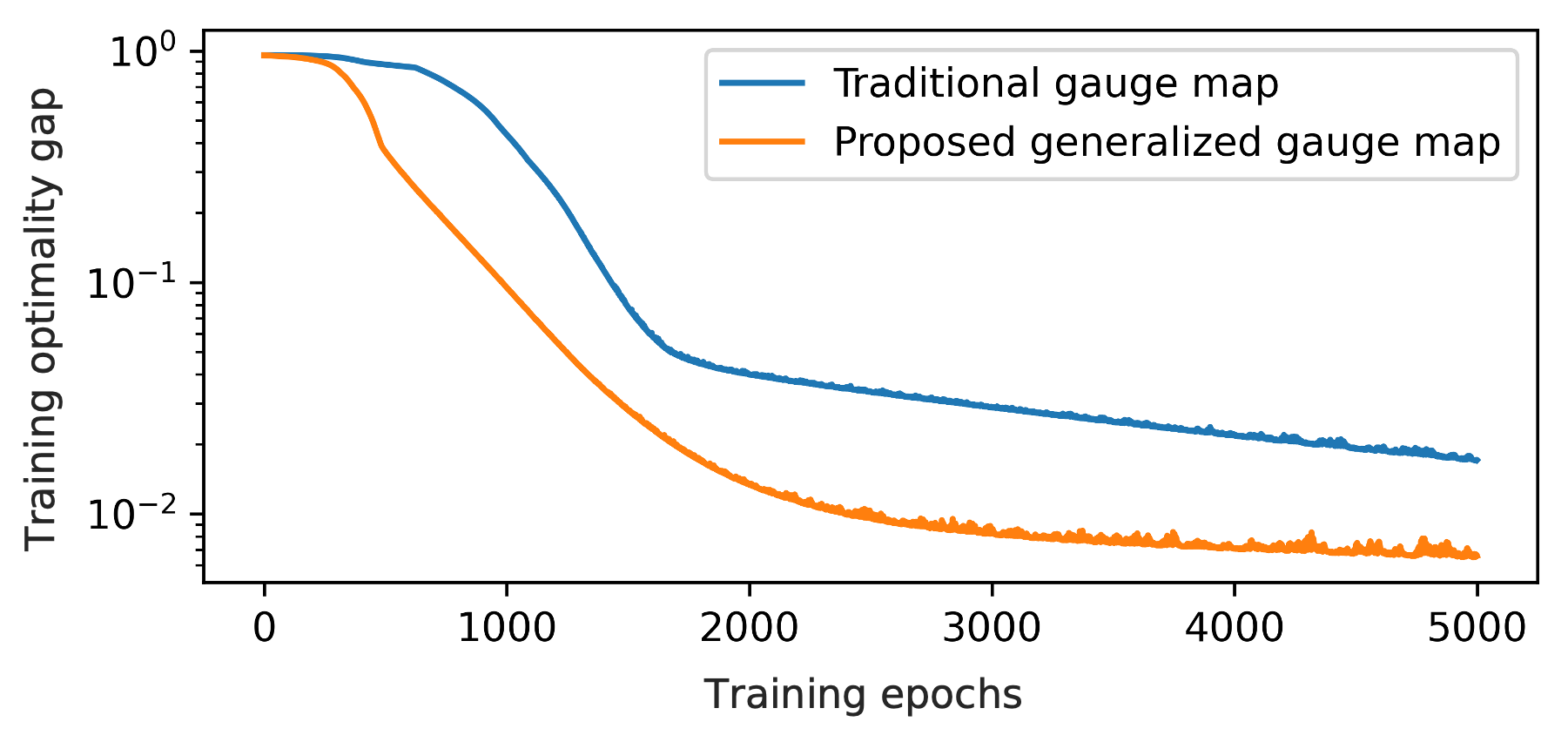}
\caption{Comparison of Training loss (optimality gap) using traditional gauge map as in \eqref{eq:T} and the proposed generalized gauge map. The proposed generalized gauge map shows stronger searching speed and faster training process. } 
\label{f:train_err}
\end{figure}

\subsection{Results regarding online test}

\begin{table}[b]
\caption{Online test results using using different methods to solve the power dispatch problem. The searching time is reported as the average per instance in milliseconds. The Optimality gap is measured as $\frac{1}{N}
    \sum_{i=1}^{N}
    \left \|  \mathbf{u}^{\blacktriangle(i)}-\mathbf{u}^{(i)*} \right \|_2^2$. The Feasibility gap is calculated using $\frac{1}{N}$ $\left ( \mathbf{1}^{\texttt{T}}\max (\mathbf{A}_{\texttt{ineq}}\mathbf{u}^{\blacktriangle(i)}+\mathbf{B}_{\texttt{ineq}}\mathbf{x}^{(i)}+\mathbf{b}_{\texttt{ineq}},\mathbf{0}) +\right.$ $ \left. \mathbf{1}^{\texttt{T}}| \mathbf{A}_{\texttt{eq}}\mathbf{u}^{(i)}+\mathbf{B}_{\texttt{eq}}\mathbf{x}^{(i)}+\mathbf{b}_{\texttt{eq}} |\right )$. }\centering
\begin{tabular}{|>{\centering\arraybackslash}p{2.5cm}|>{\centering\arraybackslash}p{1.5cm}|>{\centering\arraybackslash}p{1.5cm}|>{\centering\arraybackslash}p{1.5cm}|}
\hline
Method                                    & Optimality gap & Feasibility gap & Search time (ms) \\ \hline

Penalty method \cite{liu2022teaching},  $\rho$=1e-6 & 0.012          & 0.084           & 0.052              \\ \hline
Penalty method \cite{liu2022teaching},  $\rho$=10   & 1.146          & 0.015           & 0.048              \\ \hline
Gurobi solver \cite{gurobi}                            & 0.000          & 0.000           & 16.43              \\ \hline
Projection method \cite{zhao2020deepopf+}                         & 0.013          & 0.000           & 11.77              \\ \hline
Traditional gauge \cite{li2023learning}                    & 0.019          & 0.000           & 0.100              \\ \hline
\LOOPLCP                                 & 0.011          & 0.000           & 0.055              \\ \hline
\end{tabular}
\label{rtable}
\end{table}

Table \ref{rtable} presents the results of using different methods to solve the power dispatch problem. We tuned different penaly coefficients to maximize the performance and Table \ref{rtable} only lists the results with best optimality gap and that with best feasibility
gap. Our results show that among the methods ensuring hard feasibility,  the \LOOPLCP~method surpasses other methods in terms of optimality and search time.

\section{Conclusion}
\begin{figure}[htbp]
\centering
\includegraphics[width=0.7\columnwidth]{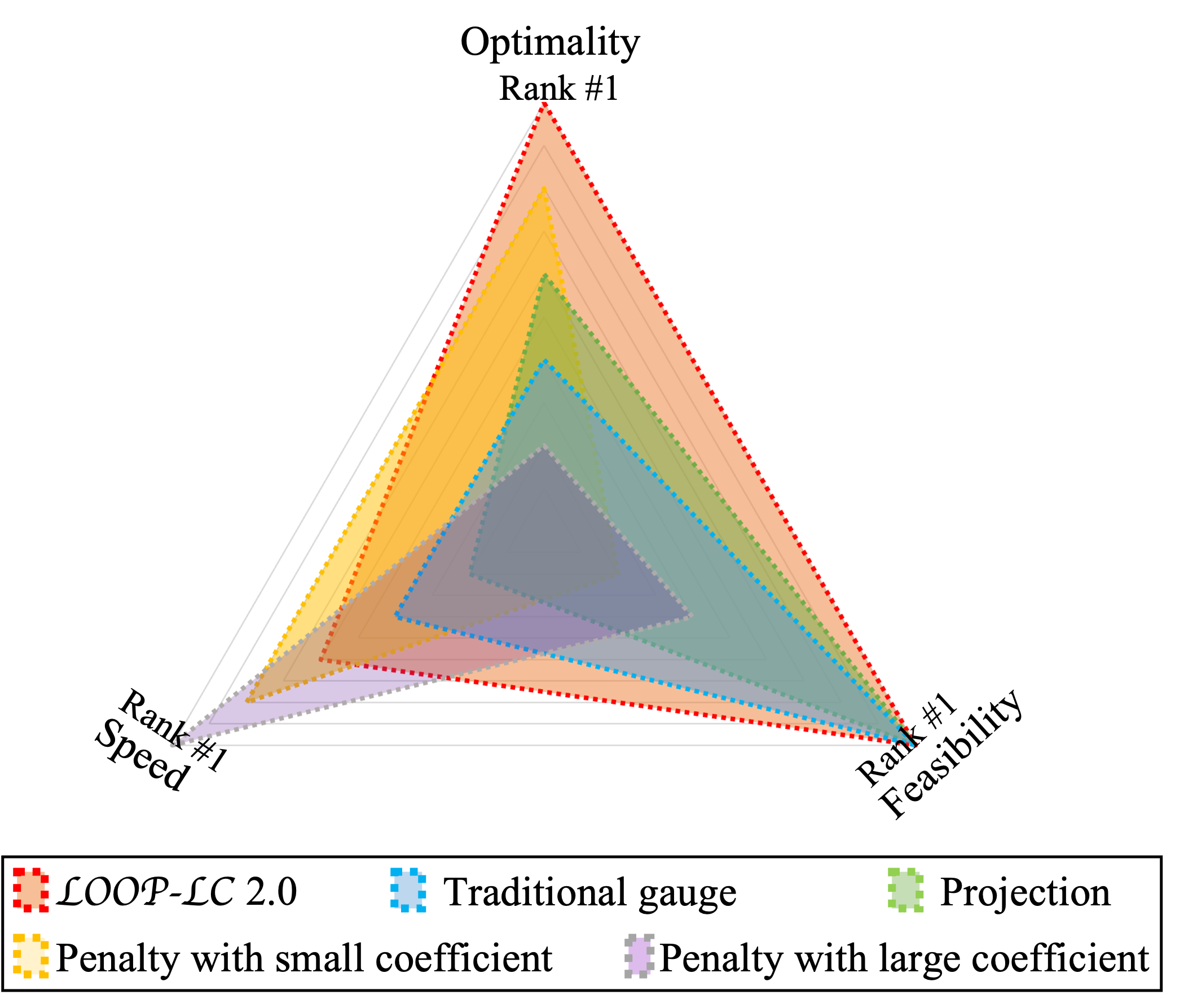}
\caption{Radar chart comparing the ranking of different ML-driven methods in optimality, feasibility, and speed. The proposed \LOOPLCP~ranks top in all metrics. The best performance in each dimension is noted as Rank\#1.} 
\label{radar}
\end{figure}

This paper introduces the \LOOPLCP~ model for solving power dispatch problems with hard linear constraints. At its core, we propose a generalized gauge map to transfer infeasible solutions to the feasible range.  Unlike current learning-assisted solutions, our
method is free of parameter tuning and removes iterations altogether. Fig.\ref{radar} shows the radar chart of different
methods. Among the
methods ensuring hard feasibility, the \LOOPLCP~ method surpasses other methods in optimality and search time.

\vspace{-.1cm}
\bibliographystyle{ieeetr}
\bibliography{main.bib}

\end{document}